# A Trio-Method for Retinal Vessel Segmentation using Image Processing

Mahendra Kumar Gourisaria[1][0000-0002-1785-8586] Vinayak Singh[1][0000-0003-3052-588x] Manoj Sahni[2][0000-0001-7949-9147]

[1] School of Computer Engineering, KIIT Deemed to be University, Bhubaneswar - 751024, Odisha, India

[2] Department of Mathematics, Pandit Deendayal Energy University, Gandhinagar 382426, Gujarat, India

`mkgourisaria2010@gmail.com`

**Abstract.** Inner Retinal neurons are a most essential part of the retina and they are supplied with blood via retinal vessels. This paper primarily focuses on the segmentation of retinal vessels using a triple preprocessing approach. DRIVE database was taken into consideration and preprocessed by Gabor Filtering, Gaussian Blur, and Edge Detection by Sobel and Pruning. Segmentation was driven out by 2 proposed U-Net architectures. Both the architectures were compared in terms of all the standard performance metrics. Preprocessing generated varied interesting results which impacted the results shown by the UNet architectures for segmentation. This real-time deployment can help in the efficient preprocessing of images with better segmentation and detection.

## 1 Introduction

The retina is an essential part of Homo sapiens. These retinal vessels are highly sensitive. Retinal vessel segmentation is a crucial process for better diagnosis and analysis of the retina. Manual analysis of retinal vessels is very time consuming for doctors and physicians. To overcome this problem various researchers are contributing to the segmentation of retinal vessels for the ease of physicians [1]. However, as per previous observations, it was noticed that results gained while retinal vessels segmentation were not much satisfactory with respect to our result [2]. Most of the research articles were lacking in preprocessing and deep analysis of the performance of the segmentation model by using various performance metrics [3-5]. To overcome this problem, we have used the trio-pre-processing approach namely Gabor Filtering, Gaussian Blur, and Edge Detection by Sobel and Pruning for retinal vessel segmentation by using our proposed two U-Net Architecture. Most of the researchers concentrated on AUC, Accuracy, and Sensitivity but in our work, we have taken in to consideration with all the metrics, especially with IoU score, efficacy score, and accuracy.

## 2 Materials and Methods

### 2.1 Dataset Used and Augmentation

Dataset was collected from the Grand Challenge portal [6]. The dataset consisted of a total of 40 images and masks respectively where image size was resized from (565 x 584) pixels to (512 x 512) pixels. As per the dataset, the training set was small so image augmentation plays an important role. In our approach, horizontal flip, vertical flip, and rotation were considered for generating the augmented images and masks. In terms of dataset splitting, 80 images for training, 20 images for testing, and 20 images for validation were considered.

### 2.2 Proposed Methodology

For better results, three approaches were considered for training and testing of the proposed U-Net Model: Approach 1 consisted of Gaussian Blur, Approach 2 was based on Gabor Filtering, whereas Approach 3 was based on edge detection by Sobel and Pruning. Fig. 1 shows the images after image processing.

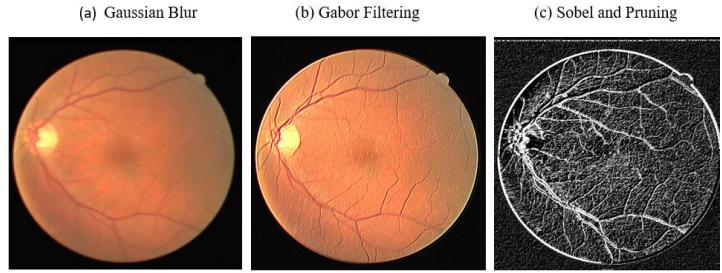

(a) Gaussian Blur  (b) Gabor Filtering  (c) Sobel and Pruning

Fig. 1 (a) shows the processed image using Gaussian Blur, (b) shows the processed image using Gabor Filtering, and (c) shows the processed image from Sobel and Pruning

In total two U-Net architectures namely: Reti-UNet1 and Reti-UNet2 were implemented for the segmentation of retinal vessels. Reti-UNet1 consisted of 4 encoders and 4 decoders and Reti-UNet1 consisted of 5 encoders and 5 decoders respectively. Table 1 shows the configuration of U-Net models. All the architectures were trained on 100 epochs with a batch size of 2 and the learning rate was set to 0.0001.

Table 1. Configuration of U-Net Models

| Model | Encoder Layers | Bridge Layer | Decoder Layers |
| --- | --- | --- | --- |
| Reti-UNet1 | {64,128,256,512} | [512,1024] | {512,256,128,64} |
| Reti-UNet2 | {64,128,256,512,512} | [512,1024] | {512,512,256,128,64} |

## 3 Results and Discussion

For evaluation, various performance metrics like IoU Score (IS), Dice Coefficient (DC), Dice Loss (DL), Accuracy (Acc.), Recall (Rec.), Training Time (TT), and Efficacy Ratio (ER) were considered for finding out the best and optimal U-Net model with low computational cost. Eq. (1) shows the formula for the Efficacy ratio:

$$\text{Efficacy Ratio} = \text{IoU Score} / \text{Dice Loss} \tag{1}$$

Table 2, 3, and 4 shows the performance metrics of all approaches with the proposed U-Net model. From table 2, we can observe that Reti-UNet1 performed outstandingly when compared with all other architecture. Reti-UNet1 with Gaussian Blur performed excellently well by gaining the highest IOU score of 0.7708, validation accuracy of 0.9671 efficacy score of 1.9748 which is very high. This architecture can provide us with good segmentation results at a low computational cost. Fig. 2 shows the output predicted by Reti-UNet-1 by using Gaussian Blur.

Table 2. Performance Metrics with Approach 1 (Gaussian Blur)

| Model | IS | Acc. | Rec. | DL | DC | TT (in seconds) | ER |
|---|---|---|---|---|---|---|---|
| **Reti-UNet1** | **0.7708** | **0.9671`** | **0.7547** | **0.3903** | **0.7263** | **3968** | **1.9748** |
| Reti-UNet2 | 0.7668 | 0.9659 | 0.7461 | 0.3895 | 0.7570 | 4039 | 1.9686 |

Table 3. Performance Metrics with Approach 2 (Gabor Filtering)

| Model | IS | Acc. | Rec. | DL | DC | TT (in seconds) | ER |
|---|---|---|---|---|---|---|---|
| Reti-UNet1 | 0.7637 | 0.9652 | 0.7334 | 0.3969 | 0.7506 | 3603 | 1.9241 |
| Reti-UNet2 | 0.7660 | 0.9659 | 0.7378 | 0.3902 | 0.7544 | 3620 | 1.9630 |

Table 4. Performance Metrics with Approach 3 (Edge Detection by Sobel and Pruning)

| Model | IS | Acc. | Rec. | DL | DC | TT (in seconds) | ER |
|---|---|---|---|---|---|---|---|
| Reti-UNet1 | 0.7519 | 0.9622 | 0.7170 | 0.4273 | 0.7287 | 3514 | 1.7596 |
| Reti-UNet2 | 0.7465 | 0.9617 | 0.7047 | 0.4365 | 0.7296 | 3597 | 1.7101 |

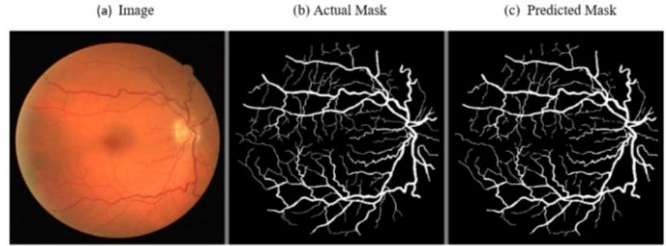

Fig. 2 Prediction by Reti-UNet-1 by using Gaussian Blur on the test set

## 4 Conclusion and Future Work

The proposed workflow using Gaussian Blur gained the highest accuracy. Our study emphasizes providing an efficient Reti-UNet model for retinal vessel segmentation. We have compared all the architectures based on various performance metrics. Table 5 shows the comparison of our work with other previous related works. Our best-proposed Reti-UNet1 by using Gaussian Blur performed excellently by gaining the IoU score of 0.7708, Efficacy ratio of 1.9748, and Accuracy of 0.9671. In terms of image processing, we had three types of image processing where Gabor Filtering was not appropriate for the dataset but Gaussian Blur performed very well among all the image processing techniques. In terms of future work, attention UNet models and pre-trained models can be used for retinal vessel segmentation. Various image processing techniques can be used for enhancing the vessels for better segmentation and detection.

Table 5. Comparison of our work with Related Work

| Author | Khan *et al.* [2] | Dong *et al.* [3] | Wang *et al.* [4] | Yang et al. [5] | **Our work** |
|---|---|---|---|---|---|
| Accuracy | 0.9610 | 0.9586 | 0.9567 | 0.9500 | **0.9671** |